%% file: v3.4.tex
\documentclass[prl,a4,nofootinbib,floatfix,twocolumn,showpacs,superscriptaddress]{revtex4}
\usepackage[dvips]{pstcol}
\usepackage{graphicx,amsmath}
\usepackage{dcolumn}
\usepackage{bm}
\usepackage{rotating}
\usepackage{amsmath,amssymb,amsfonts}
\usepackage{color}	
\usepackage{epsfig}

\def\hermes{{\sc Hermes~}}

\def\desy{{\sc Desy~}}
\def\hera{{\sc Hera~}}

\def\pythia6{{\sc Pythia6~}}

\def\jetset{{\sc Jetset~}}

\newcommand{\ms}{\mskip 1.5mu}
\newcommand{\Ms}{\mskip 2.0mu}
\newcommand{\bs}{\mskip -1.5mu}

\newcommand{\im}{\operatorname{Im}}

\newcommand{\ru}[4]{u^{\ms #1 #2}_{#3 #4}}

\newcommand{\rs}[4]{s^{\ms #1 #2}_{#3 #4}}
\newcommand{\rn}[4]{n^{\ms #1 #2}_{#3 #4}}
\newcommand{\rN}[4]{N^{\ms #1 #2}_{\Ms #3 #4}}
\newcommand{\rU}[4]{U^{\ms #1 #2}_{\Ms #3 #4}}
\newcommand{\0}{{\mskip 2.5mu} 0 {\mskip 2.5mu}}

\begin{document}


\title{Exclusive \boldmath{$\rho^0$} electroproduction on transversely polarized protons\\
      }

\input{authors3}

\date{\today}

\begin{abstract}
The exclusive electroproduction of $\rho^0$ mesons was studied with the
\hermes spectrometer at the \desy laboratory by scattering 27.6 GeV positron
and electron beams off a transversely polarized hydrogen target.
Spin density matrix elements for this process were determined from the measured 
production- and decay-angle distributions of the produced $\rho^0$ mesons.
These matrix elements embody information on helicity transfer and
the validity of $s$-channel helicity conservation in the case of
a transversely polarized target.
From the spin density matrix elements, the leading-twist term in the single-spin
asymmetry was calculated separately for longitudinally and transversely
polarized $\rho^0$ mesons.
Neglecting $s$-channel helicity changing matrix elements, results for the former
can be compared to calculations based on generalized parton distributions,
which are sensitive to the contribution of the total angular momentum
of the quarks to the proton spin.
\end{abstract}


\maketitle


Exclusive electroproduction of mesons can provide new information
about the structure of the nucleon because of its relation to generalized
parton distributions (GPDs)~\cite{Mue,Rad,Ji}.
In Ref.~\cite{CFS} it has been proven that the amplitude for hard exclusive
electroproduction of mesons by longitudinal virtual photons can be factorized
into a hard-scattering part and a soft part
that depends on the structure of the nucleon and the produced meson.
In the case of exclusive vector meson production, also 
the produced meson is longitudinally polarized (in addition to the virtual
photon being longitudinal).
The amplitude for the soft part can be expressed in terms of GPDs.
	
GPDs provide a three-dimensional representation of the structure of the nucleon
at the partonic level, correlating the longitudinal momentum fraction of a
parton with its transverse spatial coordinates.
They are related to the standard parton distribution functions
and nucleon form factors~\cite{Ji,GPV,Diehl,BelRad}.
At leading twist, meson production is described by four types of GPDs:
$H^{q,g}$, $E^{q,g}$, $\widetilde{H}^{q,g}$, and $\widetilde{E}^{q,g}$, where $q$ stands
for a quark flavour and $g$ for a gluon.
The GPDs are functions of $t$, $x$, and $\xi$, where $t$ is
the squared four-momentum transfer to the nucleon, $x$ the average, and $\xi$
half the difference of the longitudinal momentum fractions
of the quark or gluon in the initial and final state.
The quantum numbers of the produced meson determine the sensitivity to the various GPDs.
In particular, at leading twist, production of vector mesons is sensitive only
to the GPDs $H^q, E^q, H^g$, and $E^g$.

\begin{figure}[!hb]
 \begin{center}
  \begin{minipage}{8.0cm}
   \begin{center}
     \epsfig{file=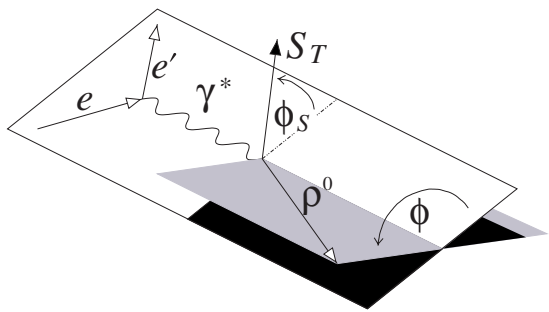,width=0.9\textwidth,angle=0}
        \end{center}
     \caption{The lepton scattering and hadron production planes 
         together with the azimuthal angles $\phi$ and $\phi_S$.}
   \label{fig:phiphis}
  \end{minipage}
 \end{center}
\end{figure}

The transverse target-spin asymmetry in exclusive electroproduction of
longitudinally polarized vector mesons by longitudinal virtual photons
is an important observable, because it depends almost linearly
on the GPD $E$~\cite{GPV}. 
This is in contrast to the unpolarized cross section, where the contribution of $E$ is
generally small compared to the contribution of $H$.
At leading twist, the asymmetry is proportional to $\sin (\phi-\phi_S)$,
where $\phi$ and $\phi_S$ are the azimuthal angles about the virtual-photon
direction of the hadron production plane and the transverse part $\vec{S}_T$
of the target spin, respectively, with respect to the lepton scattering plane 
(see Fig.~\ref{fig:phiphis}). 

\begin{figure}[!htb]
 \begin{center}
  \begin{minipage}{7.0cm}
   \vspace{0.3cm}
    \begin{center} \tiny \hspace{-0.3cm}
      \epsfig{file=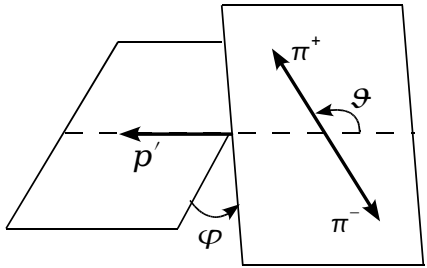,width=0.5\textwidth,angle=-90}
     \end{center}
   \end{minipage}
 \end{center}
 \caption{The polar and azimuthal angles of the decay $\pi^+$ of the
          $\rho^0$ in the $\rho^0$ rest frame.
          The positive $z$-axis is taken opposite to the direction
          of the residual proton, while the angle $\varphi$ is defined
          with respect to the hadron production plane.}
\label{fig:rhodec}
\end{figure}

The cross section and asymmetry for exclusive $\rho^0$ electroproduction $ e + p \rightarrow e' + \rho^0 + p'$ can conveniently be described using spin density matrix elements~\cite{WS,Fra,D}. 
By using the angular distribution of the produced vector meson and of its decay
products, as described by the polar and azimuthal angles $\vartheta$ and
$\varphi$ (see Fig.~\ref{fig:rhodec}), one can separate the contributions
of mesons with longitudinal and transverse polarization to the measured asymmetries. 
If $s$-channel helicity conservation (SCHC) holds, the helicity of the
virtual photon is transferred to the produced vector meson. 
In that case studying the asymmetry for the production of longitudinally
polarized vector mesons is tantamount to selecting longitudinal virtual photons.
Measurements have shown that SCHC holds reasonably well for exclusive electroproduction
of $\rho^0$ mesons on an unpolarized target at \hermes kinematics~\cite{SDME}.	
Thus information on the GPD $E$ can be obtained from measurements of the
transverse target-spin asymmetry in exclusive $\rho^0$ electroproduction.
\newline
Ultimately, these studies will help to understand the origin of the nucleon spin,
because it has been shown~\cite{Ji} that the $x$-moment in the limit
$t \rightarrow 0$ of the sum of the GPDs $H^q$ and $E^q$
is related to the contribution $J^q$ of the total angular momentum
of the quark with flavour $q$ to the nucleon spin.
	
In this paper, measurements of exclusive $\rho^0$ electroproduction on 
transversely polarized protons are presented. For the first time,
values of the spin density matrix elements (SDMEs) and the transverse
target-spin asymmetry for this process were determined.

The data were collected with the \hermes spectrometer~\cite{hermes:spectr} 
during the period $2002 - 2005$. The 27.6 GeV \hera electron or positron beam
at \desy scattered off a transversely polarized hydrogen target~\cite{target}
of which the spin direction was reversed every $1 - 3$ minutes.
The average magnitude of the target polarization was $|P_T| = 0.724 \pm 0.059$.
The lepton beam was longitudinally polarized, the helicity being reversed
periodically. The net polarization for the selected data was $0.095 \pm 0.005$,
mainly because more data were taken with positive helicity. 

Leptons were distinguished from hadrons with an average efficiency of $98\%$
and a hadron contamination of less than $1\%$ by using the information from
an electromagnetic calorimeter, a transition-radiation detector, a preshower 
scintillation counter, and a Ring Imaging \v{C}erenkov detector. 
Events were selected in which only one lepton and two oppositely charged hadrons
were detected.

In the event selection, the following kinematic constraints were imposed:
$Q^2>1$ GeV$^2$, $W^2>4$ GeV$^2$, and $-t' < 0.4$ GeV$^2$.
Here $-Q^2$ is the squared four-momentum of the exchanged virtual photon,
$W$ the invariant mass of the virtual-photon proton system, and
$t'$ the reduced Mandelstam variable $t'=t-t_0$, where $-t_0$ is the
minimum value of $-t$ for a given value of $Q^2$ and  the Bjorken variable $x_B$.
The average value of $W^2$ for the exclusive $\rho^0$ sample was 25~GeV$^2$.
The condition on $t'$ was applied to reduce non-exclusive background.


\begin{figure}[!bt]
 \begin{center}
  \begin{minipage}{8.5cm}
   \begin{center}
      \epsfig{file=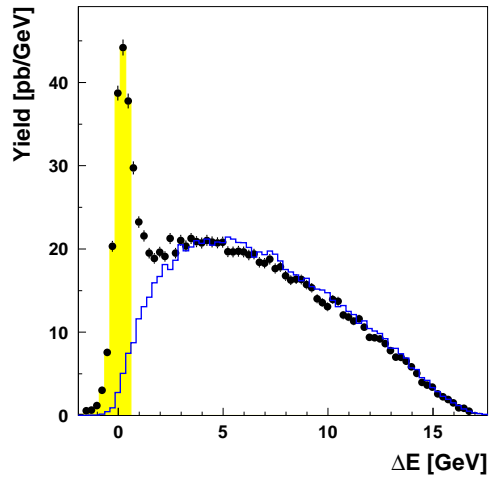,width=0.8\textwidth,angle=-0}
   \end{center} \vspace{-0.8cm}
      \caption{
         The $\Delta E$ distributions of the measured yield (number of counts
         within the acceptance divided by the integrated luminosity) (dots) and
         a Monte Carlo simulation with \pythia6\ of the non-exclusive background
         normalized to the same integrated luminosity (histogram).
         The kinematic cuts and the requirements $0.6$ GeV $< M_{\pi\pi} < 1.0$ GeV
         and $M_{KK} > 1.04$ GeV were applied.
         The selected exclusive region is indicated by the dashed area.
             }\label{fig:dedist}
    \end{minipage}
  \end{center}
\end{figure}

An exclusive event sample was selected by constraining the value of
the variable 
\begin{equation}
\Delta E = \frac{M^2_X - M^2 }{2 M} ,
\end{equation}
where $M_X$ is the missing mass and $M$ the proton mass. The measured
$\Delta E$ distribution, which includes constraints on the invariant mass of
the produced hadron pair as discussed below, is shown in Fig.~\ref{fig:dedist}. 
The peak around zero originates from the exclusive reaction. Exclusive
events were selected by the requirement $\Delta E < 0.6$ GeV.
This resulted in a total number of 7488 events.
The background from non-exclusive processes in the exclusive region was
estimated by using a \pythia6\ Monte Carlo simulation~\cite{PYTHIA6,liebing}
in conjunction with a special set of \jetset fragmentation parameters,
tuned to provide an accurate description of deep-inelastic hadron
production in the \hermes kinematic domain~\cite{hillenb, bino}.
The simulation gave a very good description of the $\Delta E$ distribution
in the non-exclusive region.
The background fractions in the exclusive region varied between 7\% and 23\%,
depending on the value of $Q^2$, $x_B$, or $t'$,
with an average over all selected data of 11\%.

The invariant mass of the two-hadron system $M_{\pi\pi}$ was determined
assuming that both hadrons are pions.
Resonant $\pi^+ \pi^-$ pairs, i.e., pairs produced in the decay
$\rho^0 \to \pi^+ \pi^-$, were selected by the condition
$0.6$ GeV $< M_{\pi\pi} < 1.0$ GeV.
Contributions in the $M_{\pi\pi}$ spectrum from the decay of a 
$\phi$ meson into two kaons were excluded by requiring
$M_{KK} > 1.04$ GeV, where $M_{KK}$ is the invariant mass of the two-hadron 
system calculated assuming that both hadrons are kaons.
After subtracting the simulated contribution from the non-exclusive tail in the
region $\Delta E < 0.6$ GeV and correcting for the non-constant acceptance with
$M_{\pi\pi}$, the $M_{\pi\pi}$ spectrum for exclusive events was fitted with a
$\rho^0$-peak plus a linear background. For the shape of the $\rho^0$-peak 
S\"oding and Ross-Stodolsky parametrizations were used. In both cases the
resulting background was found to be negligible $(0.7 \pm 0.5)\%$.


In the analysis the recently developed formalism for electroproduction
of a vector meson from a polarized nucleon was used~\cite{D}.
The cross section for exclusive $\rho^0$ leptoproduction is written as
\begin{align}
  \label{xsec1}
 \frac{d\sigma}{d\psi\, d\phi\, d\varphi\, d(\cos\vartheta)\,
                 dx_B\, dQ^2\, dt}  =   \nonumber   \\
  \frac{1}{(2\pi)^2}\, \frac{d\sigma}{dx_B\, dQ^2\, dt}\,
    W(x_B,Q^2,t,\phi,\phi_S,\varphi,\vartheta),
\end{align}
with $\psi$ being a similar angle as $\phi_S$, but now defined around the direction
of the lepton beam, and
\begin{equation}
  \label{xsec2}
\frac{d\sigma}{dx_B\, dQ^2\, dt} =
   \Gamma_v \biggl( \frac{d\sigma_T}{dt} 
         + \varepsilon\, \frac{d\sigma_L}{dt} \biggr) ,
\end{equation}
where $\Gamma_v$ is the virtual photon flux factor in the Hand convention~\cite{Han63},
$\varepsilon$ is the virtual-photon polarization parameter, and $d\sigma_T/dt$
and $d\sigma_L/dt$ are the usual $x_B, Q^2$, and $t$ dependent $\gamma^* p$
cross sections for transverse and longitudinal virtual photons, respectively.

The function $W(x_B,Q^2,t,\phi,\phi_S,\varphi,\vartheta)$ describes the
angular distribution of both the produced $\rho^0$ and its decay pions.
It consists of several terms corresponding to different polarizations
of the incoming lepton beam and the target nucleon:
\vspace{-0.1cm}
\begin{eqnarray}
\label{Wtot}
W = W_{UU} + P_\ell W_{LU} &+& S_L W_{UL} + \nonumber \\
  P_\ell S_L W_{LL} &+& S_T W_{UT} + P_\ell S_T W_{LT},
\end{eqnarray} 
where the left (right) subscript specifies the beam (target) polarization: 
unpolarized ($U$), longitudinally ($L$), or transversely ($T$) polarized,
and $P_\ell$, $S_L$, and $S_T$ represent the longitudinal polarization of the
beam, and the longitudinal and transverse polarization of the target
(with respect to the virtual photon direction), respectively.

For the case of zero beam polarization and only transverse target
polarization\footnote{
Because the target polarization is transverse to the
incoming beam, there is a small longitudinal polarization with respect
to the direction of the virtual photon. The effect of the latter
and of a small longitudinal polarization of the beam will be discussed later.}
the angular-distribution function reads
\vspace{-0.1cm}
\begin{equation}
\label{Wtrans}
W(\phi,\phi_S,\varphi,\vartheta) = W_{UU}(\phi,\varphi,\vartheta)
          + S_T W_{UT}(\phi,\phi_S,\varphi,\vartheta) \, .
\end{equation} 
Here and in the following the dependence of the various angular distribution functions
$W$ on $x_B, Q^2$, and $t$ is omitted for the sake of legibility.

The functions $W_{UY}$ (with $Y = U,T$) can be further decomposed into terms
corresponding to specific $\rho^0$ polarizations, indicated by the superscripts, according to
\begin{align}
  \label{WUY}
 W_{UY}(\phi_S,\phi,\varphi,\vartheta)  = \frac{3}{4\pi} \biggl[\,
     \cos^2\bs\vartheta\; W_{UY}^{LL}(\phi_S,\phi) + \nonumber \\
     \sqrt{2} \cos\vartheta\, \sin\vartheta\;
                          W_{UY}^{LT}(\phi_S,\phi,\varphi)
   + \sin^2\bs\vartheta\; W_{UY}^{TT}(\phi_S,\phi,\varphi)
   \,\biggr] .
\end{align}
Note that in the case of $W_{UU}$ there is no dependence on $\phi_S$. 
The production of a longitudinally polarized $\rho^0$ is described
by $W_{UY}^{LL}$, the production of a transversely polarized $\rho^0$
(including the interference from amplitudes with positive and negative
$\rho^0$ helicity) by $W_{UY}^{TT}$, while $W_{UY}^{LT}$ results from the
interference between longitudinal and transverse $\rho^0$ polarizations. 

The terms $W_{UY}^{AB}$ can be expanded (see Eqs. 4.10 and 4.17 of Ref.~\cite{D})
into trigonometric functions of the angles $\phi_S,\phi$, and $\varphi$,
where the coefficients are SDMEs (or combinations thereof)
$\ru{\nu}{\nu'}{\mu}{\mu'}$ for $W_{UU}^{AB}$, and
$\rn{\nu}{\nu'}{\mu}{\mu'}$ and $\rs{\nu}{\nu'}{\mu}{\mu'}$ for $W_{UT}^{AB}$.
Here the letters $u, n$, and $s$ stand for unpolarized, normal, and sideways 
(with respect to the direction of the virtual photon and the electron scattering plane)
target polarization,
and the sub(super)scripts refer to the helicity of the virtual photon ($\rho^0$ meson)
in the helicity amplitudes that occur in the SDMEs.
In the case of $W_{UU}$ there are 15 independent terms in the expansion.
There is a direct relation between these SDMEs and the ones in
the Schilling-Wolf formalism~\cite{WS}.
For $W_{UT}$ the expansion contains 30 independent terms.


First the 15 `unpolarized' SDMEs of $W_{UU}$ were determined by fitting the
angular distributions of the combined events for the two target polarization states.
The fit was performed by maximum-likelihood estimation with 
a probability density function 
\begin{equation} \label{pdfUU}
          f_U(\phi,\varphi,\vartheta) = \mathcal{N}_U^{-1} \,
	  	\mathcal{A}(\phi,\varphi,\vartheta) \, W_{UU}(\phi,\varphi,\vartheta) ,
\end{equation}
where the function $\mathcal{A}$ represents the acceptance of the
\hermes spectrometer.
The factor $\mathcal{N}_U$ represents the normalization integral of the
probability density function, which was computed numerically using
Monte Carlo events that are within the acceptance of the spectrometer.
The non-exclusive background was included in the fit function using
fixed effective values of the SDMEs for this background.
The latter were obtained from a fit of the angular distribution of
the \pythia6\ Monte Carlo events for $\Delta E < 0.6 $~GeV.
The results for the 15 unpolarized SDMEs, which as mentioned are for
data taken in the years $2002-2005$, are fully consistent with those
from the analysis of all data taken in the period $1996-2005$
using the Schilling-Wolf formalism~\cite{SDME}.

\begin{figure*}
        \begin{center}
          \includegraphics[width=0.8\textwidth]{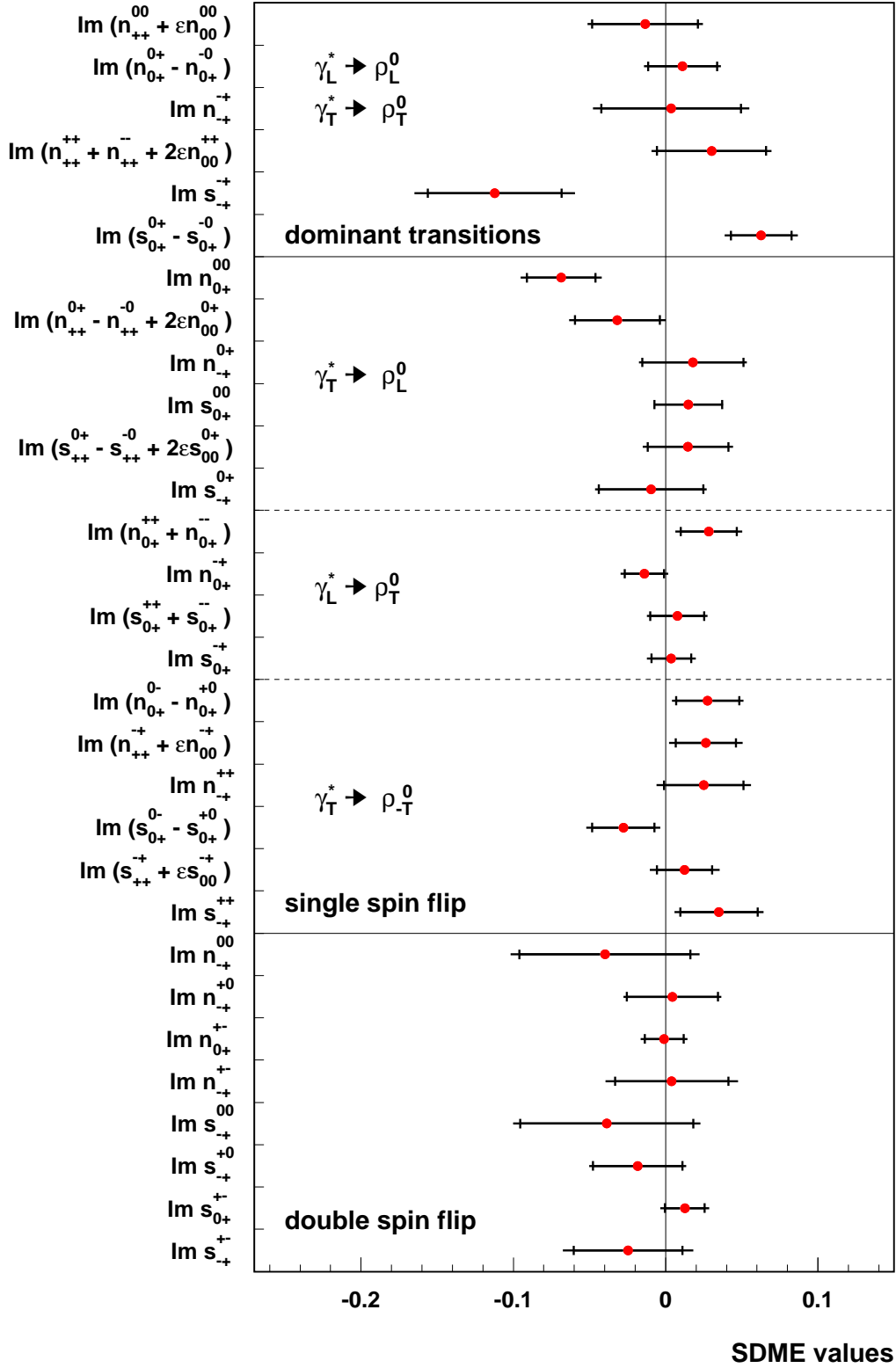}
                \caption{
Values of SDMEs, or combinations thereof, for a transversely polarized proton
target and an unpolarized beam.  The SDMEs are sorted into three categories,
which are separated from each other by the solid horizontal lines.
From top to bottom:
SDMEs containing $s$-channel helicity-conserving amplitudes,
combinations containing at least one $s$-channel helicity-changing amplitude,
and SDMEs containing two $s$-channel helicity-changing amplitudes.
Within the second category the combinations are sorted into three groups
associated with different virtual photon and $\rho^0$ polarizations.
The inner error bars represent the statistical uncertainties.
The full error bars represent the quadratic sum of the statistical
and systematic uncertainties.
In addition there is an overall scale uncertainty of 8.1\% due to the
uncertainty in the target polarization.
                }
		\label{sdmestpol}
        \end{center}
\end{figure*}

Then the 30 SDMEs $W_{UT}$ were determined, keeping the unpolarized SDMEs fixed
to the values found in the fit of $W_{UU}$ described above, using
the probability density function 
\begin{eqnarray} \label{pdfUT}
	f_T(\phi,\phi_S,\varphi,\vartheta) & = & \mathcal{N}_T^{-1} \, 
	\mathcal{A}(\phi, \phi_S,\varphi,\vartheta) \times
	\nonumber \\
	 && \big( W_{UU}(\phi,\varphi,\vartheta) + P_T W_{UT}(\phi,\phi_S,\varphi,\vartheta) \big) .  \nonumber \\
         && 
\end{eqnarray}
A factor ${d\psi}/{d\phi_S}$, which takes into account that the yields are evaluated
differentially in the angle $\phi_S$, rather than in $\psi$, was left out,
since it was very close to unity.
As  in the unpolarized case, the background was included in the fit. 
Since nothing is known about the asymmetry of the background,
the 30 SDMEs for the background were taken to be zero, and the
possible influence of this assumption was included in the systematic
uncertainties.

Besides the target polarization, various other sources of systematic uncertainties for
the SDMEs and asymmetries extracted were investigated and evaluated.
In most cases the resulting systematic uncertainties were found to be negligible,
i.e., very small compared to the statistical uncertainty.
These include the effect of radiative corrections and the uncertainties
resulting from the uncertainty in the unpolarized SDMEs and the background fraction.
The uncertainty due to the angular dependence and asymmetry of the background
was taken as the difference between a fit with a background with no angular dependence
and asymmetry, and one having the same angular dependence and asymmetry as the data.
The resulting uncertainty was found to be negligible.

\begin{figure}
        \begin{center}
        \begin{minipage}{8.5cm}
           \begin{center}
                \epsfig{file=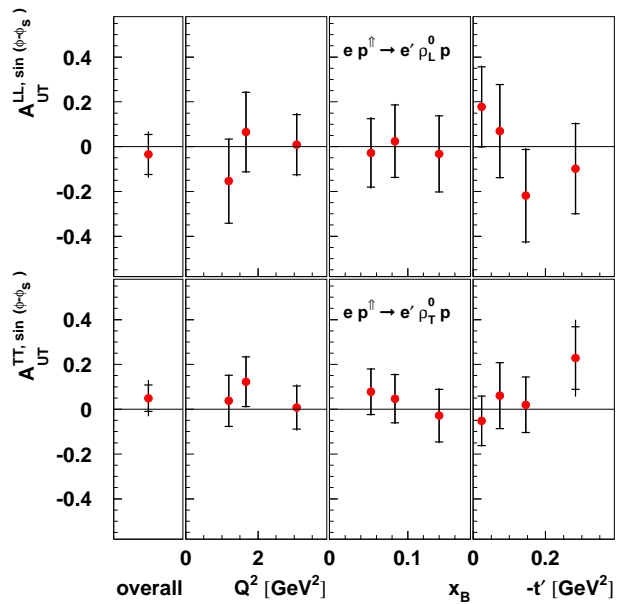,width=0.99\textwidth}
\caption{
The extracted amplitudes of the $\sin(\phi-\phi_S)$ component of $A_{UT}$ for
longitudinally (top) and transversely polarized (bottom) $\rho^0$ mesons.
The inner error bars represent the statistical uncertainties. The full error 
bars represent the quadratic sum of the statistical and systematic uncertainties.
In addition there is an overall scale uncertainty of 8.1\% from the uncertainty in the target polarization.
         }
		\label{fig:DDS}
           \end{center}
        \end{minipage}
        \end{center}
\end{figure}

The influence of the net beam polarization of approximately 0.095 was estimated
by including the SDMEs for $W_{LU}$ and $W_{LT}$ in the fit.
Even if the latter had large uncertainties, the influence on the ones
for $W_{UT}$ was negligible.
The data presented in Fig.~\ref{fig:DDS} are effectively integrated over all or two
of the variables $Q^2$, $x_B$, and $t'$ within the experimental acceptance.
The effect of this kinematic averaging was estimated by comparing the results of a
Monte Carlo simulation that included a modelled dependence of the asymmetry on
these variables with the model input values at the average kinematics.
Also this effect was found to be negligible.

In the extraction of the SDMEs the small longitudinal component of the target
polarization with respect to the direction of the virtual photon
(the average value of $|S_L/P_T|$ was 0.072) was neglected.
This component introduces a term $S_L W_{UL}$, which is described by 14 SDMEs.
As the value of $S_L$ is small, these SDMEs cannot be determined from the present data.
A systematic uncertainty was estimated by using several sets of random values
obeying the positivity bounds given in Ref.~\cite{D} for these SDMEs,
and evaluating the resultant changes.
Changes of on average 55\% of the statistical uncertainty were found,
with a maximum of 76\% for one SDME 
($\im(\rs{-}{+}{+}{+} + \epsilon \rs{-}{+}{\0}{\0}$)).
This is the main source of systematic uncertainty. 

Lastly there are systematic uncertainties arising from misalignment
of the detector, detector smearing effects, and bending of the beam and produced
charged particles in the transverse holding field of the target magnet.
The uncertainties due to all effects together were investigated with a Monte Carlo
simulation of the possible influence of these effects.
The resultant uncertainty was found to be negligible. 

The resulting SDMEs are shown in Fig.~\ref{sdmestpol}.
Almost all of them are consistent with zero within $1.5 \sigma$,
where $\sigma$ represents the total uncertainty in the value of an SDME.
Note that these include s-channel helicity conserving SDMEs.
Similar SDMEs in the unpolarized case were found ~\cite{SDME}
to be non-zero and large (0.4 - 0.5).
The SDMEs $\im\bigl( \rs{\0}{+}{\0}{+} - \rs{-}{\0}{\0}{+} \bigr)$,
$\im\rs{-}{+}{-}{+}$, and $\im\rn{\0}{\0}{\0}{+}$ deviate more than 
$2.5 \sigma$ from zero.
The former two involve the interference between natural ($N$)
and unnatural ($U$) parity exchange amplitudes~\cite{D}.
For instance, $\im\rs{\0}{+}{\0}{+}$ contains
the product $\rN{\0}{+}{\0}{+}  (\rU{+}{+}{+}{-})^*$
and $\im\rs{-}{+}{-}{+}$ contains
the product $\rN{-}{+}{-}{+}  (\rU{+}{+}{+}{-})^*$.
The detailed analysis of unpolarized data has shown that
$\rN{\0}{+}{\0}{+}$  and $\rN{-}{+}{-}{+}$ are dominant $N$ amplitudes.
The $U$ amplitudes presumably are small, as they are suppressed at large $Q^2$.
However, $\rU{+}{+}{+}{-}$ is relatively large~\cite{SDME,Man}.
The SDME $\im\rn{\0}{\0}{\0}{+}$ corresponds to a $\gamma^*_T \rightarrow \rho_L$
transition, the SDMEs of which were found to be non-zero in the unpolarized case.
The value of $-0.069 \pm 0.022$ measured for $\im\rn{\0}{\0}{\0}{+}$
is another indication of violation of SCHC in the $\gamma^*_T \rightarrow \rho_L$ transition.


As mentioned, the $\sin(\phi-\phi_S)$ term in the transverse target-spin 
asymmetry for production of longitudinally polarized $\rho^0$ mesons
is of special importance because of its sensitivity to the GPD $E$.
The amplitude of this term is given in terms of SDMEs as~\cite{D}
\begin{equation}
\vspace{-0.1cm}
\label{AUTL}
  A_{UT}^{LL,\sin(\phi-\phi_S)} =
  \frac{\im \bigl( \rn{\0}{\0}{+}{+} + \varepsilon \rn{\0}{\0}{\0}{\0} \bigr)}
       {\ru{\0}{\0}{+}{+} + \varepsilon \ru{\0}{\0}{\0}{\0}} .
\end{equation}
The resultant values for all selected data and for bins in $x$, $Q^2$,
and $t'$ are shown in Fig.~\ref{fig:DDS}~(top).
They are all zero within the error bars.
Because the SCHC violating terms $\im(\rn{\0}{\0}{+}{+})$ and $\ru{\0}{\0}{+}{+}$
in Eq.~\ref{AUTL} require a double helicity flip (see Ref.~\cite{D} for details),
they presumably can be neglected.
Then the value of $A^{LL,\sin(\phi-\phi_S)}_{UT} = -0.035 \pm 0.103$
\footnote{This is the value for 'all' data, which has average kinematics
$<Q^2>=1.95$ GeV$^2$, $<x_B>=0.08$, and $<-t>'=0.13$ GeV$^2$.}
can be compared to the results of GPD calculations for the production of a longitudinally
polarized $\rho^0$ by a longitudinal photon $A^{\sin(\phi-\phi_S)}_{UT,\gamma^*_L,\rho_L}$,
which is given by $\im(\rn{\0}{\0}{\0}{\0}) /\ru{\0}{\0}{\0}{\0}$.

The $\sin(\phi-\phi_S)$ amplitude for the production of transversely polarized
$\rho^0$ mesons is given by
\begin{equation}
\vspace{-0.05cm}
\label{AUTT}
  A_{UT}^{TT,\sin(\phi-\phi_S)} =
  \frac{\im\bigl( \rn{+}{+}{+}{+} + \rn{-}{-}{+}{+} + 2\varepsilon \rn{+}{+}{\0}{\0} \bigr)}
       {1-(\ru{\0}{\0}{+}{+} + \varepsilon \ru{\0}{\0}{\0}{\0})}.
\vspace{-0.05cm}
\end{equation}
The values for this asymmetry are shown in Fig.~\ref{fig:DDS}~(bottom).
Also these are zero within the error bars.

A few groups have performed GPD-based calculations of the transverse target
asymmetry for exclusive $\rho^0$ production.
In Refs.~\cite{GPV,vinnikov} the quark GPD $E^q$ is parametrized in terms of
the value of $J^u$, taking $J^d=0$.
Ref.~\cite{vinnikov} includes the contribution of gluons.
The calculated values of $A^{\sin(\phi-\phi_S)}_{UT,\gamma^*_L,\rho_L}$
are in the range 0.15 to 0.00 for $J^u=$ 0.0 to 0.4.
In Refs.~\cite{DieKu,GolKr} the GPDs are modeled using data for nucleon form
factors, sum rules and positivity constraints.
The results of both calculations are similar. Values of $J^u$ and $J^d$
of approximately 0.22 and 0.00, respectively, are found, and the calculated values
of the asymmetry are very small ($-0.03$ to 0.02), which is consistent with the
present data.
It must be realized that the results depend on the modelling of the relevant
GPDs of both quarks and gluons, and that the kinematic conditions of the
calculations are in several cases outside the kinematic range of the present data.

In summary, the transverse target single-spin asymmetry was measured for
exclusive $\rho^0$ electroproduction on a transversely polarized hydrogen
target. Spin density matrix elements were determined by using the angular
distributions of the produced $\rho^0$ mesons and their decay into two pions.
Almost all of the SDMEs describing transverse target polarization
were found to be consistent with zero. A notable exception is an SDME that 
corresponds to the production of a longitudinally polarized $\rho^0$ by a
transverse virtual photon. The fact that it is non-zero indicates a small
violation of $s$-channel helicity conservation in the case of a transversely
polarized target.
The amplitude of the $\sin(\phi-\phi_S)$ component of the asymmetry for the production
of longitudinally polarized $\rho^0$ mesons was found to be small ($-0.035 \pm 0.103$).
Neglecting double helicity changing SDMEs, this component can be identified with the
leading-twist term of the asymmetry.
Calculations based on generalized parton distributions predict
small values, consistent with the measured value.

\begin{acknowledgments}  
We thank M. Diehl for providing us with the theoretical formalism in an early stage and for many interesting and helpful discussions.
We gratefully acknowledge the \desy\ management for its support and the staff at \desy\ and the collaborating institutions for their significant effort.
This work was supported by the FWO-Flanders, Belgium;
the Natural Sciences and Engineering Research Council of Canada;
the National Natural Science Foundation of China;
the Alexander von Humboldt Stiftung;
the German Bundesministerium f\"ur Bildung und Forschung (BMBF);
the Deutsche Forschungsgemeinschaft (DFG);
the Italian Istituto Nazionale di Fisica Nucleare (INFN);
the MEXT, JSPS, and COE21 of Japan;
the Dutch Foundation for Fundamenteel Onderzoek der Materie (FOM);
the U. K. Engineering and Physical Sciences Research Council, the
Particle Physics and Astronomy Research Council and the
Scottish Universities Physics Alliance;
the U. S. Department of Energy (DOE) and the National Science Foundation (NSF);
the Russian Academy of Science and the Russian Federal Agency for 
Science and Innovations;
the Ministry of Trade and Economical Development and the Ministry
of Education and Science of Armenia;
and the European Community-Research Infrastructure Activity under the
FP6 ''Structuring the European Research Area'' program
(HadronPhysics, contract number RII3-CT-2004-506078).

\end{acknowledgments}
\vspace{-0.3cm}


\end{document}

%% file: authors3.tex

\def\groupargonne{\affiliation{Physics Division, Argonne National Laboratory, Argonne, Illinois 60439-4843, USA}}
\def\groupbari{\affiliation{Istituto Nazionale di Fisica Nucleare, Sezione di Bari, 70124 Bari, Italy}}
\def\groupbeijing{\affiliation{School of Physics, Peking University, Beijing 100871, China}}
\def\groupcolorado{\affiliation{Nuclear Physics Laboratory, University of Colorado, Boulder, Colorado 80309-0390, USA}}
\def\groupdesy{\affiliation{DESY, 22603 Hamburg, Germany}}
\def\groupzeuthen{\affiliation{DESY, 15738 Zeuthen, Germany}}
\def\groupdubna{\affiliation{Joint Institute for Nuclear Research, 141980 Dubna, Russia}}
\def\grouperlangen{\affiliation{Physikalisches Institut, Universit\"at Erlangen-N\"urnberg, 91058 Erlangen, Germany}}
\def\groupferrara{\affiliation{Istituto Nazionale di Fisica Nucleare, Sezione di Ferrara and Dipartimento di Fisica, Universit\`a di Ferrara, 44100 Ferrara, Italy}}
\def\groupfrascati{\affiliation{Istituto Nazionale di Fisica Nucleare, Laboratori Nazionali di Frascati, 00044 Frascati, Italy}}
\def\groupgent{\affiliation{Department of Subatomic and Radiation Physics, University of Gent, 9000 Gent, Belgium}}
\def\groupgiessen{\affiliation{Physikalisches Institut, Universit\"at Gie{\ss}en, 35392 Gie{\ss}en, Germany}}
\def\groupglasgow{\affiliation{Department of Physics and Astronomy, University of Glasgow, Glasgow G12 8QQ, United Kingdom}}
\def\groupillinois{\affiliation{Department of Physics, University of Illinois, Urbana, Illinois 61801-3080, USA}}
\def\groupmichigan{\affiliation{Randall Laboratory of Physics, University of Michigan, Ann Arbor, Michigan 48109-1040, USA }}
\def\groupmoscow{\affiliation{Lebedev Physical Institute, 117924 Moscow, Russia}}
\def\groupnikhef{\affiliation{National Institute for Subatomic Physics (Nikhef), 1009 DB Amsterdam, The Netherlands}}
\def\groupstpetersburg{\affiliation{Petersburg Nuclear Physics Institute, Gatchina, Leningrad region, 188300 Russia}}
\def\groupprotvino{\affiliation{Institute for High Energy Physics, Protvino, Moscow region, 142281 Russia}}
\def\groupregensburg{\affiliation{Institut f\"ur Theoretische Physik, Universit\"at Regensburg, 93040 Regensburg, Germany}}
\def\grouprome{\affiliation{Istituto Nazionale di Fisica Nucleare, Sezione Roma 1, Gruppo Sanit\`a and Physics Laboratory, Istituto Superiore di Sanit\`a, 00161 Roma, Italy}}
\def\grouptriumf{\affiliation{TRIUMF, Vancouver, British Columbia V6T 2A3, Canada}}
\def\grouptokyo{\affiliation{Department of Physics, Tokyo Institute of Technology, Tokyo 152, Japan}}
\def\groupamsterdam{\affiliation{Department of Physics and Astronomy, Vrije Universiteit, 1081 HV Amsterdam, The Netherlands}}
\def\groupwarsaw{\affiliation{Andrzej Soltan Institute for Nuclear Studies, 00-689 Warsaw, Poland}}
\def\groupyerevan{\affiliation{Yerevan Physics Institute, 375036 Yerevan, Armenia}}
\def\groupnone{\noaffiliation}


\groupargonne
\groupbari
\groupbeijing
\groupcolorado
\groupdesy
\groupzeuthen
\groupdubna
\grouperlangen
\groupferrara
\groupfrascati
\groupgent
\groupgiessen
\groupglasgow
\groupillinois
\groupmichigan
\groupmoscow
\groupnikhef
\groupstpetersburg
\groupprotvino
\groupregensburg
\grouprome
\grouptriumf
\grouptokyo
\groupamsterdam
\groupwarsaw
\groupyerevan


\author{A.~Airapetian} \groupgiessen \groupmichigan
\author{N.~Akopov}  \groupyerevan
\author{Z.~Akopov}  \groupyerevan
\author{E.C.~Aschenauer}  \groupzeuthen
\author{W.~Augustyniak}  \groupwarsaw
\author{A.~Avetissian}  \groupyerevan
\author{E.~Avetisyan}  \groupdesy
\author{B.~Ball}  \groupmichigan
\author{N.~Bianchi}  \groupfrascati
\author{H.P.~Blok}  \groupnikhef \groupamsterdam
% ...

\collaboration{The HERMES Collaboration} \noaffiliation